\documentclass[11pt,a4paper]{iopart}
\usepackage[utf8]{inputenc}
\expandafter\let\csname equation*\endcsname\relax

\expandafter\let\csname endequation*\endcsname\relax 
\usepackage{amsmath}
\usepackage{amsfonts}
\usepackage{amssymb}
\usepackage{bm}
\usepackage{graphicx}
\usepackage{iopams}
\usepackage{todonotes}
\usepackage{hyperref}
\usepackage[normalem]{ulem} 
\usepackage{soul}

\usepackage[utf8]{inputenc}
\usepackage{amsmath}
\usepackage{amsfonts}
\usepackage{amssymb}
\usepackage{makeidx}
\usepackage{graphicx}
\usepackage{color}
\usepackage{slashbox}
\usepackage{pifont}

\usepackage{xcolor}

\usepackage{bm}
\def\bea{\begin{eqnarray}}
\def\eea{\end{eqnarray}}
\def\ba{\begin{array}}
\def\ea{\end{array}}

\def\la{\langle}
\def\ra{\rangle}

\usepackage[utf8]{inputenc}
\usepackage{amsmath}
\usepackage{amsfonts}
\usepackage{amssymb}
\usepackage{hyperref}

\newcommand{\opunit}{\textrm{1}\kern-0.22em\textrm{l}}

\def\bea{\begin{eqnarray}}
\def\eea{\end{eqnarray}}
\def\ba{\begin{array}}
\def\ea{\end{array}}

\def\la{\langle}
\def\ra{\rangle}

\begin{document}

\title{Forces from coarse-graining nonequilibrium degrees of freedom: exact results}
\author{Ion Santra and Matthias Kr\"uger}
\address{Institute for Theoretical Physics, University of G\"ottingen, 37077, G\"ottingen, Germany}

\begin{abstract}
We explore the relaxation dynamics of a tracer in a harmonic trap coupled to a non-equilibrium bath particle in stationary state, finding qualitative differences compared to the well known equilibrium case.  These can be attributed to an additional position and time dependent force acting on the tracer, emerging when averaging the bath degree in the non-equilibrium stationary state conditioned to a certain tracer position. Specifically, we provide analytical results for 
an overdamped tracer coupled linearly to a bath particle in different nonequilibrium scenarios, namely, subjected to a different temperature than the tracer, or to active noise with Gaussian or non-Gaussian fluctuations. For the  case of different temperatures, the conditioned tracer-bath force can be as large in magnitude as the force from the trapping potential. For an active bath particle with memory,  even the bath noise takes a finite average under conditioning of the tracer. Further, if the noise of the bath particle is non-Gaussian, the relaxation function of the tracer can be non-monotonic as a function of time. We also compute the \emph{pinned relaxation},  proposing that measurement and comparison of conditioned and pinned relaxation allows determination of the non-equilibrium forces in experiments.

\end{abstract}

\maketitle
\section{Introduction}

The Langevin equation is the paradigmatic  stochastic description of Brownian motion, proposed by Paul Langevin in 1908~\cite{langevin1908}.
It explains Browian motion in Newtonian fluids. For complex fluids like viscoelastic systems, where the fluid's relaxation time is comparable to that of the Brownian particle, a generalized Langevin equation (GLE) is used. This GLE incorporates memory, with dissipation depending on the tracer's history and including colored noise~\cite{zwanzigbook}, as e.g., found in the independent oscillator model developed by Feynmann-Vernon~\cite{feynman2000theory,caldeira1983path}, or the Rubin model of a chain of oscillators~\cite{rubin1961statistical}. For equilibrium, such procedure appears well understood,  with the noise following the fluctuation dissipation theorem (FDT)~\cite{kubo1966fluctuation}, and the Zwanzig Mori projection operator formalism  presenting a formal derivation of such GLEs for the tracer degree of freedom starting from  Liouville's equations~\cite{mori1965continued,mori1965transport,zwanzigbook}. In derivations of GLEs for linear systems~\cite{feynman2000theory,caldeira1983path,rubin1961statistical,panja2010generalized,goychuk2012viscoelastic},   the bath degrees are averaged over the conditioned equilibrium distribution of the bath 
for a given initial tracer position \cite{zwanzig1973nonlinear,bez1980microscopic,hanggi2007} so that additional terms, called ``initial slip" \cite{hanggi2007}, vanish.

Out of equilibrium \cite{maes2015friction,steffenoni2016interacting,maggi2017memory,kruger2016modified, berner2018oscillating, demery2019driven,te2019mori,schilling2022coarse,netz2023derivation,santra2023dynamical,sarkar2024harmonic}, however, 
the stationary  conditioned distributions for bath degrees are not known in general, and neither is their influence on the tracer dynamics. In this paper, we analyze paradigmatic models  
of a tracer particle coupled to an overdamped  bath particle \cite{goychuk2012viscoelastic,siegle2010markovian,maes2020fluctuating,jain2021two,santra2023dynamical}.
Such models 
have been found to capture many aspects of experiments on Brownian particles in viscoelastic surroundings
\cite{muller2020properties,jain2021two,ginot2022barrier,caspers2023mobility}. Specifically, we study an overdamped tracer trapped in a harmonic trap and linearly coupled to a single overdamped non-equilibrium bath particle, this linear system allowing for analytical analysis. The bath particle is thereby subject to different types of non-equilibrium noises: In case (i) tracer and bath particles are maintained at different temperatures. Case (ii) considers an active Gaussian bath, i.e., an active Ornstein-Uhlenbeck particle. Case (iii) is an active non-Gaussian bath, where the bath particle follows one-dimensional run-and-tumble dynamics.

We derive the stationary distributions of bath degrees conditioned to a given tracer position for these cases. Using them, we observe that these non-equilibrium cases show a intriguing extra forces in the resulting GLE, which, in contrast to equilibrium, do not vanish in the stationary average, and which influence the tracer's relaxation dynamics in a qualitative way. As an example observed in case (i), the initial slope of the relaxation function is influenced by the bath, in contrast to equilibrium. In case (ii), an additional intricate force occurs in the GLE, which is due to the stationary distribution of the noise of the bath particle. Finally in case (iii), these forces can render the relaxation function to be non-monotonic as a function of time.

We finally demonstrate that a so called pinned relaxation, where the tracer is pinned at a fixed position and then released to relax, is unaffected by the mentioned non-equilibrium phenomena, i.e., it follows the same GLE but without the mentioned extra forces. We thus propose to use the difference between pinned and stationary relaxation as a means to  experimentally detect and investigate these forces.

We describe the model and the general mathematical setup in Sec.~\ref{s:model}; we present the well-known equilibrium result in Sec.~\ref{s:eq}. Then we move to the nonequilibrium scenarios: in Sec.~\ref{s:thneq} we study the case where the tracer and bath are maintained at different temperatures; in Sec.~\ref{s:aoup} and Sec.~\ref{s:rtp} we investigate the cases of a Gaussian and a non-Gaussian active baths, respectively. Finally, we conclude in Sec.~\ref{s:concl}.




\begin{figure}
\centering\includegraphics[width=.8\hsize]{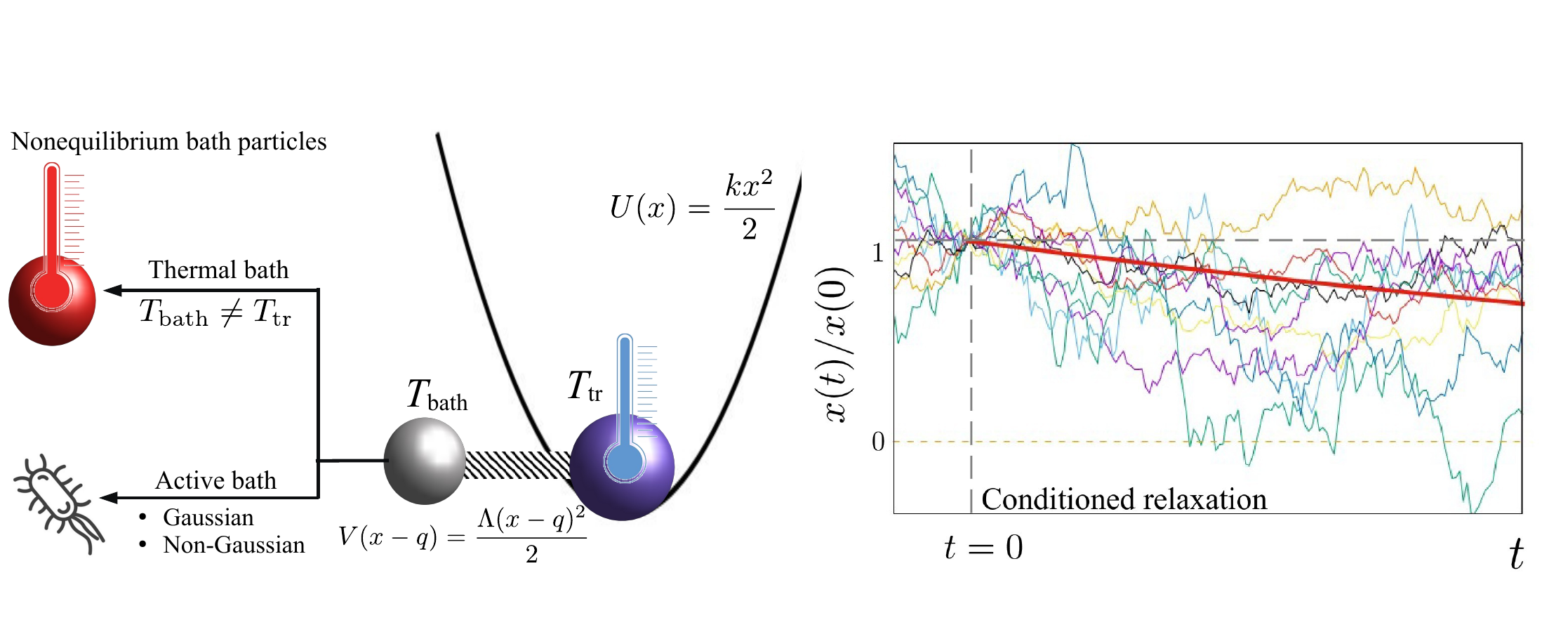}
\caption{Left: Schematic representation of the setup: Tracer particle (blue, position $x$), subject to harmonic trap $U(x)=kx^2/2$, and coupled to a bath particle (grey, position $q$) via a harmonic interacting potential $V(x-q)$. Arrows on the left indicate the different scenarios considered. The right panel shows example trajectories of the relaxation function from position $x_0$ in the stationary state, obtained by simulating Eq.~\eqref{scaled:model:quad} with $T_{\text{bath}}/T_{\text{tr}}=10$, $\gamma=1$, $k=1$, and $\Lambda=0.1$, while thick red line represents the analytical result obtained for the relaxation function for these parameters.}
\label{f:model}
\end{figure}

\section{Tracer-Bath Model: Introduction and general results}\label{s:model}
In this section, we specify the studied tracer bath model \cite{goychuk2012viscoelastic,muller2020properties} and discuss its general solutions. These are then, in the following sections, specified according to the nature of the bath.

Consider an overdamped tracer particle, with friction coefficient $\gamma_\text{tr}$, coupled to an overdamped bath particle with coefficient $\gamma_{\text{bath}}$ via a harmonic potential with spring coefficient $\bar\Lambda$. The tracer is kept in a harmonic potential with coefficient $k$ and is subject to noise at  temperature $T_\text{tr}$. 
We introduce dimensionless equations, where lengths are written in  units  of $\sqrt{k_\text{B}T_{\text{tr}}/k}$, time in units of $\gamma_\text{tr}/k$, and energy in  units of $k_{\text{B}}T_{\text{tr}}$. In these units, the dimensionless positions $x$ and $q$ of tracer and bath, respectively, follow the Langevin equations, with $\gamma=\gamma_\text{bath}/\gamma_{\text{tr}}$ and $\Lambda=\bar\Lambda /k$,
\begin{subequations}
\begin{align}
\dot{x}(t)&=-x-\Lambda(x-q)+\eta(t)\label{eom:tracer1}\\
\gamma\dot{q}(t)&=-\Lambda(q-x)+f(t)\label{eom:bath1}.
\end{align}\label{scaled:model:quad}
\end{subequations}
The Gaussian noise $\eta(t)$ is characterized by $\la \eta(t)\ra=0$, and $\la \eta(t)\eta(t')\ra=2\delta(t-t')$; $f(t)$, the noise of the bath particle, is a stationary stochastic process with zero mean, with other properties specified for the given cases in the respective sections below.  Due to the presence of the harmonic trap, the average position of the tracer in the stationary state is zero. How does the tracer relax to this value?  

To answer this question, one notes that, since the equations of motion Eq.~\eqref{scaled:model:quad} are linear in $(x,q)$,  one can integrate out the bath position to obtain an exact Langevin equation for the time evolution of the tracer,
\begin{align}
\int_{0}^{t}\Gamma(t-t')\dot{x}(t')dt'=-x(t)+ F(t).\label{equi:lle}
\end{align}
 The  kernel $\Gamma(t)$ and noise $F(t)$ are given by (introducing $\nu\equiv\frac{\Lambda}{\gamma}$),
\begin{align}
\Gamma(t)&=\delta(t)+\Lambda e^{-\nu t},\label{eff:diss:eq}\\
F(t)&=\eta(t)+\nu\int_{0}^{t} dt' e^{-\nu(t-t')}f(t')+\Lambda(q_0-x_0)e^{-\nu t}\label{eff:noise:eq}.
\end{align}
The noise $F$ depends on the initial positions of the tracer and the bath, $x_0$ and $q_0$. We are interested in finding the relaxation of the position of the tracer for $t\geq 0$ from its position $x_0$ at $t=0$. To obtain it, we average  Eq.~\eqref{equi:lle} under the condition $x_0$ at time $t=0$, denoting that average $\langle \dots \rangle_{x_0}$. This requires the corresponding averages of the stochastic processes $f$ and $q$.  
The mean of the noise $F(t)$ for $t\geq 0$ under  this condition is thus 
\begin{align}
    \la F(t) \ra_{x_0}&=\Lambda\la(q_0-x_0)\ra_{x_0}e^{-\nu t}+\nu\int_{0}^{t} dt' e^{-\nu(t-t')}\la f(t')\ra_{x_0}\cr
    &\equiv F^{I}(x_0,t)+F^{II}(x_0,t).\label{lin:gen}
\end{align}
The second equality  defines the forces $F^I$ and $F^{II}$, which are still to be computed in the specific cases of the following sections. $F^I$ results from a non-vanishing initial tracer-bath force, and $F^{II}$ is the consequence of a finite bath noise $f$ under the condition.

The relaxation function, i.e., $\mathcal{R}(x_0,t)=\la x(t)\ra_{x_0}$, can now be found from the generalized Langevin equation, Eq.~\eqref{equi:lle}, which, in general, yields an inhomogeneous differential equation for $\mathcal{R}(x,t)$, 
\begin{align}
      \int_{0}^tdt'\Gamma(t-t')\dot{\mathcal{R}}(x,t')&=-\mathcal{R}(x,t)+F^I(x,t)+F^{II}(x,t).\label{aoup:rxt:1}
 \end{align}
  Noting that $\mathcal{R}(x,0)=x$ by definition, the solution $\mathcal{R}_P(x,t)$ of the homogeneous part of the above equation, i.e., for $F^{II}(x,t)=0$ and $F^{I}(x,t)=0$, is given by,
 \begin{align}
    \mathcal{R}_P(x,t)=x e^{-\frac{c_1t}{2}}\left[\cosh\left(\frac{c_2t}{2}\right)+\frac{c_1-2}{c_2}\sinh\left(\frac{c_2t}{2}\right)\right],\label{eq:pinned:gen}
\end{align}
 with $c_1=(1+\Lambda+\nu)$, and $c_2=\sqrt{c_1^2-4\nu}$. The physical relevance of this homogeneous solution $\mathcal{R}_P(x,t)$, giving rise to its notion of  {\it pinned relaxation} will be discussed later in Sec.~\ref{s:thneq}. One can then write the general solution of Eq.~\eqref{aoup:rxt:1} as,
 \begin{align}
     \mathcal{R}(x,t)=\mathcal{R}_P(x,t)+\int_0^t dt' \frac{\mathcal{R}_P(x,t-t')}{x}\left( F^{I}(x,t') + F^{II}(x,t') \right).\label{mcd:rtplin}
 \end{align}
If the noise $f$ decays exponentially with a time-scale $\tau$ (as is the case for the discussed scenarios), i.e., if $\la f(t)\ra_{x_0}=\la f\ra_{x_0} e^{-t/\tau}$, the time integrals in Eq.~\eqref{mcd:rtplin} can be performed, leading to, 
\begin{align}
     \mathcal{R}(x,t)=\mathcal{R}_P(x,t)+ F^{I}(x,0) h^I(t)+\left\la f\right\ra_x h^{II}(t).\label{cd:general}
 \end{align}
The functions $h^I(t)$ and $h^{II}(t)$ are given by,
\begin{align}
     h^I(t)&= 2e^{- \frac{c_1t}{2}} c_2^{-1} \sinh\left(\frac{c_2t}{2}\right)\label{hI},\\
     h^{II}(t)&=\frac{e^{- \frac{c_1t}{2}}\left[c_2^{-1}(c_1-2\tau^{-1})\sinh(c_2t/2))+\cosh(c_2t/2)\right]-e^{-t/\tau}}{(\tau^{-2}-c_1\tau^{-1}+\nu)}\label{hII},
\end{align}
with $c_{1,2}$ given below Eq.~\eqref{eq:pinned:gen}. 

What remains to be done in the following sections is to find $\langle q_0-x_0\rangle_{x_0}$ and $\left\la f\right\ra_{x_0}$ for the considered setups, and to discuss the physical relevance of $\mathcal{R}_P(x,t)$.

\section{ Equilibrium case}\label{s:eq}
Despite being well known \cite{goychuk2012viscoelastic,muller2020properties,caspers2023mobility}, we first review the equilibrium case, for a self contained presentation. Here the tracer and the bath particle are subject to noise at the same temperature, i.e., $f(t)$, in Eq.~\eqref{eom:bath1} is a Gaussian white noise with the same properties as $\eta(t)$, 
\begin{align}
\la f(t)\ra&=0\\
\la f(t)f(t')\ra&=2\gamma\delta(t-t').\label{eq:wnoise}
\end{align}
The steady state distribution is naturally given by the Boltzmann distribution, i.e., the joint distribution of $x_0$ and $q_0$ reads
\begin{align}
P(x,q)=\frac{1}{Z}e^{-\Lambda\frac{(q-x)^2}{2}}e^{-\frac{x^2}{2}} .\label{conditioned:eq}
\end{align}
This reiterates that $x_0$ and $x_0-q_0$ are, in equilibrium, statistically independent. The conditioned distribution for $q_0$ is easily found
\begin{align}
P(q|x_0)=\frac{\int dx\, \delta(x-x_0)P(x,q)}{\int dq\int dx\, \delta(x-x_0)P(x,q)}=\frac{1}{\sqrt{4\pi /\Lambda}}e^{-\Lambda\frac{(q-x_0)^2}{2}}.\label{eq:cond}
\end{align}
This leads to,
\begin{align*}
    \langle x_0-q_0 \rangle_{x_0}=0.
\end{align*}
Naturally, the white noise in  Eq.~\eqref{eq:wnoise} yields for $t>0$,
\begin{align*}
    \langle f (t) \rangle_{x_0}=0.
\end{align*}
%
Using these and  Eq.~\eqref{lin:gen}, the noise $F$ of the generalized Langevin equation for the tracer fulfils 
\begin{align}
   \la F(t)\ra_{x_0}&=0\label{eq:mean}.
 \end{align}
The above equations show that the forces defined in  Eq.~\eqref{lin:gen} vanish, i.e., $F^{I,II}(x,t)=0$.  The relaxation function is thus given by the homogeneous solution of   Eq.~\eqref{aoup:rxt:1}, given by  Eq.~\eqref{eq:pinned:gen}, i.e., 
\begin{align}
   \mathcal{R}(x_0,t)= \mathcal{R}_P(x_0,t).
\end{align}
It is worth noticing, as a reference for later cases, that, for short times, the relaxation function follows,
\begin{align}
   R(x_0,t)=x_0\left(1- t + \mathcal{O}(t^2)\right).
\end{align}
The linear order is thus not influenced by the presence of the bath particle. This is because, at time $t=0$, there is, on average, no force acting between tracer and bath, and the short time dynamics is determined by the force the tracer feels from the trap.

\section{Particles with different temperatures}\label{s:thneq}
We continue with a nonequilibrium setup: 
We keep the tracer noise at a temperature $T_\text{tr}$, while the bath particle is subject to noise at a temperature $T_\text{bath}$, with $T=T_\text{bath}/T_\text{tr}$ dimensionless, i.e., 
\begin{align}
\la f(t)\ra&=0\\
\la f(t)f(t')\ra&=2T\gamma\delta(t-t').\label{eq:noisef}
\end{align}
Such two-temperature setups have been used 
to study heat transport~\cite{rieder1967properties,dhar2008heat}, phase separation of polymers~\cite{grosberg2015nonequilibrium,grosberg2018dissipation,ilker2020phase,netz2020approach}, and rheological properties of viscoelastic baths~\cite{demery2019driven}.

  Unlike equilibrium, the stationary distribution in this case is not the Boltzmann distribution, i.e., it differs from Eq.~\eqref{conditioned:eq}. For this linear system it is a bivariate Gaussian which can be found analytically,  given explicitly in~\ref{app:neqdist}. Integrating it as in Eq.~\eqref{eq:cond} yields the conditioned distribution of $q$ 
  \begin{align}
     P(q|x)= \frac{1}{\sqrt{4\pi \sigma^2}}\exp\left[-\frac{(q-\langle q\rangle_x)^2}{2\sigma^2} \right]\label{neq:th:dist}.
  \end{align}
The lengthy expression for variance $\sigma^2$ is given in~\ref{app:neqdist}; the  mean is given by
  \begin{align}
      \langle q\rangle_x=x\frac{\Lambda(1+\gamma T)+\gamma T}{\Lambda(1+\gamma T)+\gamma }.\label{th:A}
  \end{align}
  We see that $\langle x-q\rangle_x$ is generally finite,   indicating that for the nonequilibrium setup, the bath particle is not distributed symmetrically about the tracer; This leads to a non-zero result for $F^I(x,t)$. The equilibrium result is recovered by noticing that  $\langle q\rangle_x=x$ for $T=1$, for which case $F^I(x,t)$ vanishes.
  
\begin{figure}
\centering    \includegraphics[width=\hsize]{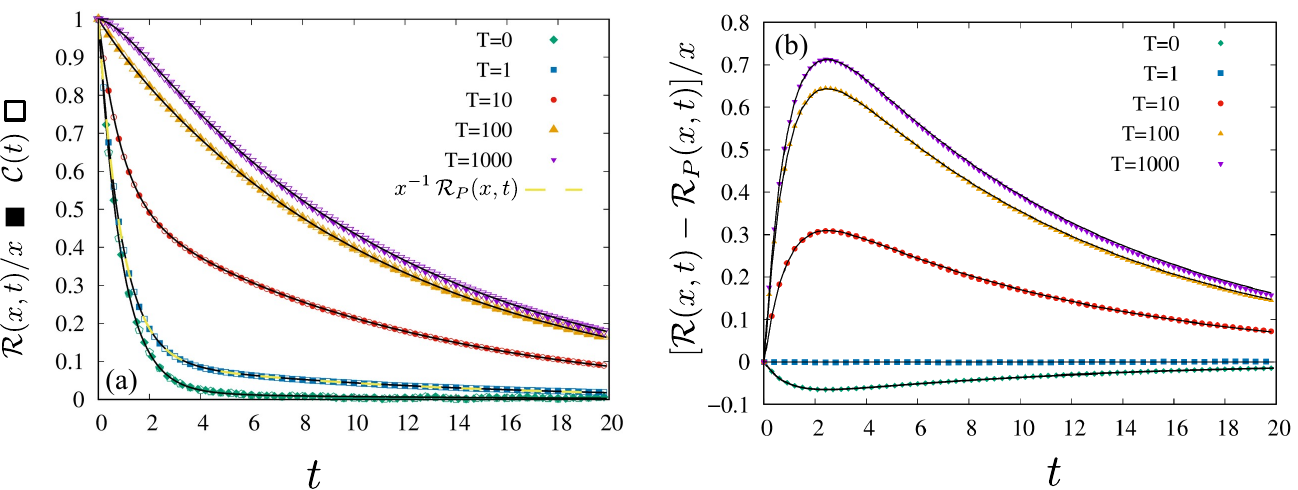}
    \caption{Different temperature case: (a) Relaxation function $\mathcal{R}(x,t)/x$ for different values of the bath temperature; Data from numerical simulations are shown as filled symbols, while solid black lines represent the analytical expression Eq.~\eqref{mcd-ac-linear:th:neq}. Pinned relaxation $\mathcal{R}_P(x,t)/x$ (yellow dashed line) is independent of bath temperature (see discussion in Fig.~\ref{pinning:scheme}). Open symbols show the position autocorrelation $\mathcal C(t)=\la x(t)x(0)\ra/\la x^2\ra$ for the same set of bath temperatures (see discussion section~\ref{autocorrelation:dis}). 
    (b)  
    Plot showing the difference between conditioned and pinned relaxation as a function of time for the same set of bath temperatures as (a). The data from numerical simulations are shown by filled symbols, while the solid black lines represent the analytical expression Eq.~\eqref{mcd-pin:neq:lin}. For both (a) and (b) $\Lambda=0.1$, and $\gamma=1$.}
    \label{fig:lin:neq:th:mcdPR}
\end{figure}

    $F^{II}(x_0,t)$ defined in Eq.~\eqref{lin:gen} is zero, since the thermal noise fulfills $\la f(t)\ra_{x_0}=0$. The conditioned mean noise $F$ using  Eq.~\eqref{th:A} is thus, 
\begin{align}
\la F(t)\ra_{x_0}=F^{I}(x_0,t)=\Lambda x_0\,\frac{\gamma(T-1)}{\gamma+\Lambda+\gamma\Lambda T}e^{-\nu t}\label{fneq}.
\end{align}
  Using this in Eq.~\eqref{mcd:rtplin}, the conditioned relaxation is found to be, 
\begin{align}
   \mathcal{R}(x,t)=\mathcal{R}_P(x,t)+x\,\frac{\Lambda\gamma(T-1)}{\gamma+\Lambda+\gamma\Lambda T} h^{I}(t)
   \label{mcd-ac-linear:th:neq}
\end{align}
where $h^{I}(t)$ is given by Eq.~\eqref{hI}. This is shown in Fig~\ref{fig:lin:neq:th:mcdPR}a), displaying agreement  with numerical simulations.

 The behavior of the conditioned relaxation at short-times can be determined by the mean force experienced by the tracer at $t=0$, i.e.,
\begin{align}
    \frac{d}{dt} \langle x(t)\rangle_{x_0}|_{t=0} =   F^I(x_0,0) -x_0.\label{eq:meanf}
    \end{align}
The above equation states that at $t=0$, in addition to the trap force $-x_0$, the particle also experiences a net force from the bath particle. It is insightful to consider the limits of large and small $T$. If $T\gg 1$, i.e., if the bath temperature is large compared to that of the tracer, \begin{align}
    \lim_{T\gg1} \frac{F^{I}(x_0,0)}{x_0}-1= -\frac{\Lambda^{-1}+\gamma^{-1}+1}{ T}.\label{highT}
\end{align}
In this limit, the net force approaches zero from below, i.e., for a very hot bath particle, the tracer is initially nearly force free. This is visible in the curves shown in Fig.~\ref{fig:lin:neq:th:mcdPR}a), where the curves for large $T$ start nearly horizontally. In the other limit, for a cold bath particle
\begin{align}
    \lim_{T\to0} \frac{F^{I}(x_0,0)}{x_0}-1= -1-\frac{1}{\Lambda^{-1}+\gamma^{-1}}.\label{lowT}
\end{align}
In this limit, the net force is larger in magnitude than in absence of the bath, which means that the tracer is, for short times, stronger driven to the origin than for the equilibrium case. 

\begin{figure}
\centering    \includegraphics[width=0.7\hsize]{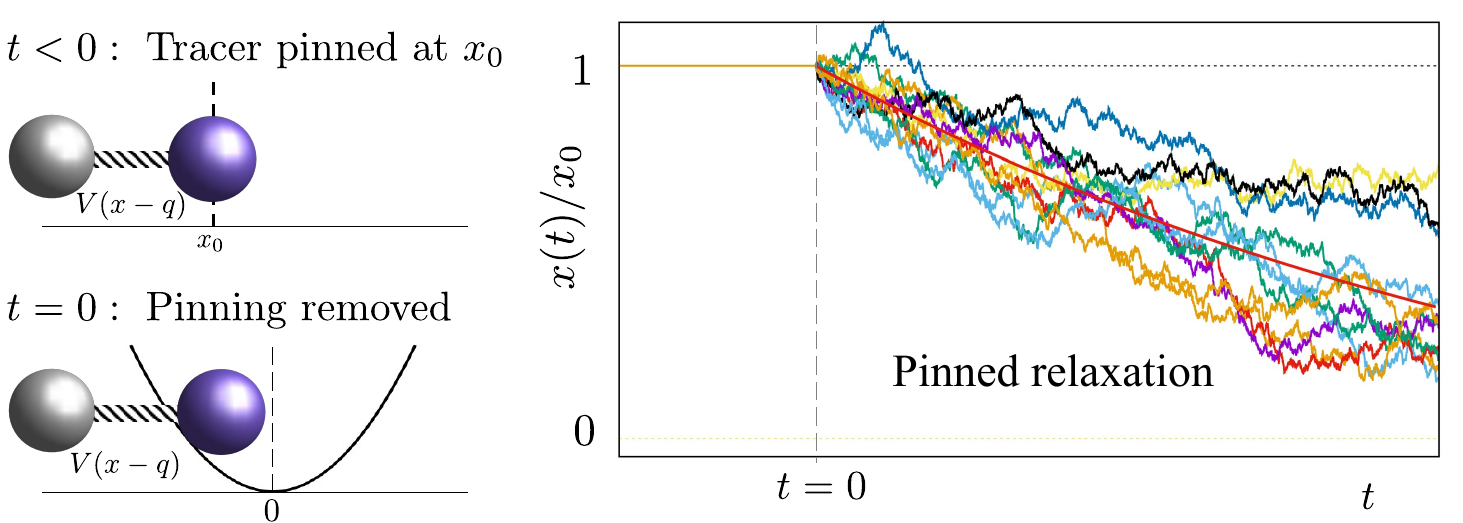}
    \caption{Schematic  of pinned relaxation. The tracer is pinned at $x_0$ for $t<0$, and the bath particle relaxes to the corresponding steady state. At $t=0$, the pinning is removed  and the tracer relaxes. Right: Sample trajectories obtained using $T_{\text{bath}}/T_{\text{tr}}=10$, $\gamma=1$, $k=1$, and $\Lambda=0.1$, with thick red line denoting the mean, i.e., $\mathcal{R}_P(x_0,t)$, see main text. }
    \label{pinning:scheme}
\end{figure}

For this case of different temperatures, the relaxation function $\mathcal{R}(x,t)$,  influenced by $F^{I}(x,t)$, differs from the homogeneous solution  $\mathcal{R}_P(x,t)$. Can 
$\mathcal{R}_P(x,t)$ be measured independently, to detect this difference? Consider therefore the pinned relaxation, Fig.~\ref{pinning:scheme}, corresponding to a protocol where the tracer is pinned (i.e., held fixed) at position $x_0$ for times $t<0$, and released at time $t=0$. Its resulting relaxation is generally different from the conditioned displacement $\mathcal{R}(x,t)$ where the tracer position $x_0$ is reached by  fluctuations. 
For the pinned relaxation, we thus compute averages at $t=0$ over the pinned distribution  $P_x(q)$. 
For a pinned tracer, the bath particle here is an equilibrium process, and $P_x(q)$ is the Boltzmann distribution at temperature $T$ 
\begin{align}
    P_{x}(q)=\frac{1}{\sqrt{4\pi T/\Lambda }}e^{-\Lambda(q-x)^2/(2T)}.\label{pinned:corr}
\end{align}
For the pinned case, we thus have
\begin{align}
    \langle x_0-q_0\rangle^P_{x_0}=0.\label{pinned:q}
\end{align}
The noise $f$ of the bath particle is not affected by the initial position, i.e.,  $\langle f\rangle^P_{x_0}=0.$
Using these, we find,
\begin{align}
    \la F(t)\ra_{x_0}^P=0
\end{align}
Thus the pinned relaxation $\mathcal{R}_P(x,t)$ evolves by,
\begin{align}
    \int_{0}^t\Gamma(t-t')\dot{\mathcal{R}_P}(x,t')&=-\mathcal{R}_P(x,t)
    \label{mpr:eom:th:lin}
\end{align}
which is  exactly the homogeneous part of Eq.~\eqref{aoup:rxt:1}. This calculation shows that the earlier found homogeneous solution of  Eq.~\eqref{aoup:rxt:1} is measured by the pinning protocol, allowing for detecting the difference between $\mathcal{R}(x,t)$ and $\mathcal{R}_P(x,t)$, given by,
\begin{align}
    \mathcal{R}(x,t)-\mathcal{R}_P(x,t)=    x_0\,\frac{\Lambda\gamma(T-1)}{\gamma+\Lambda+\gamma\Lambda T} h^I(t),
    \label{mcd-pin:neq:lin}
\end{align}
with $h^I(t)$ given by Eq.~\eqref{hI}.
The rhs of Eq.~\eqref{mcd-pin:neq:lin} starts from zero at $t=0$, reaches a maximum at $t^*$ with
\begin{align}
    t^*=\frac{2}{\sqrt{(\Lambda +\nu +1)^2-4 \nu }}\tanh^{-1}\left(\frac{\sqrt{(\Lambda +\nu +1)^2-4 \nu }}{\Lambda+\nu+1}\right),
\end{align}
 and then decreases to zero. The amplitude of Eq.~\eqref{mcd-pin:neq:lin} is a function of $T$, vanishing, as mentioned, at $T=1$, i.e, for the equilibrium case. The right hand side of Eq.~\eqref{mcd-pin:neq:lin} is shown in Fig.~\ref{fig:lin:neq:th:mcdPR}(b). We repeat that the non-zero mean of the noise $F(t)$ can be detected from the difference of pinned  and conditioned relaxations. This difference vanishes for any equilibrium bath.

\section{Gaussian bacterial bath}\label{s:aoup}
Many real world systems like living objects are inherently out of equilibrium, for example by virtue of physico-chemical processes. Such processes, popularly called `active processes' have gained immense popularity in the past decades, as evident from the huge body of works involving experiments in active microrheology with an aim to understand the tracer dynamics in active media~\cite{collectionactive1,maggi2014generalized,maggi2017memory,seyforth2022nonequilibrium,angelani2011effective,kasyap2014hydrodynamic,leptos2009dynamics,kurihara2017non,boriskovsky2024fluctuation}.  These have, in turn, triggered a number of numerical and theoretical works~\cite{morozov2014enhanced,knevzevic2020effective,burkholder2017tracer,burkholder2019fluctuation,ye2020active,maes2020fluctuating,ilker2021long,abbaspour2021enhanced,banerjee2022tracer,granek2022anomalous,abbasi2023non,shea2022passive,santra2023dynamical,asheichyk2019response}.

We thus consider in this section a model for such active particle, i.e., in the setup of  Eq.~\eqref{scaled:model:quad}, the bath particle follows an active-Ornstein-Uhlenbeck process, driven by Gaussian colored noise $f(t)$~\cite{bonilla2019active,martin2021statistical}, 
\begin{align}
\tau\dot{f}(t)&=-f(t)+\xi(t).\label{noise:aoup}
\end{align}
$\xi(t)$ is a Gaussian white noise with $\la\xi(t)\ra=0$, and $\la\xi(t)\xi(t')\ra=2\gamma T\delta(t-t')$. 
The noise $f$ driving the bath particle is thus an Ornstein Uhlenbeck process, which in the stationary state has zero mean and variance
\begin{align}
        \la f(t)f(t')\ra=\frac{\gamma T}{\tau} e^{- |t-t'|/\tau},
    \label{aoup:noisecorr}
\end{align}
The result $\int_0^\infty dt' \la f(0)f(t')\ra=\gamma T$ connects this case to Eq.~\eqref{eq:noisef}. 
Note that the variance of $f$ is not related to dissipation of the bath particle via FDT, so that this bath particle, as mentioned, is inherently out of equilibrium. The noise $f$ also adds an additional time scale $\tau$ to the problem. 

The stationary distribution is now a joint  distribution of $x$, $q$, and $f$. Due to the linearity of the system, $P(x,q,f)$ is a trivariate Gaussian,  given in~\ref{app:aoup}. The conditioned distributions of $q$ and $f$, given a tracer position $x$, can then be obtained as [see  \ref{app:aoup} for details],
\begin{align}
     P(q|x)=\frac{1}{\sqrt{4\pi \sigma_q^2}}\exp\left[-\frac{(q-\langle q\rangle_x)^2}{2\sigma_q^2} \right],\quad P(f|x)=\frac{1}{\sqrt{4\pi \sigma_f^2}}\exp\left[-\frac{(f-\langle f\rangle_x)^2}{2\sigma_f^2} \right] \label{neq:th:dist2},
  \end{align}
where the variances $\sigma^2_{q,f}$ are rather long, see \ref{app:aoup}; The mean position $\langle q\rangle_x$ reads
\begin{align}
  \langle q\rangle_x  =&x \left(1-\frac{\left[\tau  (\gamma  \Lambda +\gamma +\Lambda ) +\gamma\right]  \gamma(T-1)-\gamma  \Lambda  \tau ^2}{\Lambda  \tau ^2 (\gamma +\Lambda )+\left[\tau  (\gamma  \Lambda +\gamma +\Lambda ) +\gamma \right] (\gamma + \gamma\Lambda T+\Lambda )}\right).\label{aoup:Aq}
  \end{align}
  It is similar to  Eq.~\eqref{th:A} for the case of different temperatures, but acquires an additional term due to finite $\tau$. The force $F^{I}(x,t)$ is thus found to be
\begin{align}
    F^{I}(x,t)&=\Lambda x\frac{\left[\tau  (\gamma  \Lambda +\gamma +\Lambda ) +\gamma\right]  \gamma(T-1)-\gamma  \Lambda  \tau ^2}{\Lambda  \tau ^2 (\gamma +\Lambda )+\left[\tau  (\gamma  \Lambda +\gamma +\Lambda ) +\gamma \right] (\gamma + \gamma\Lambda T+\Lambda )}e^{-\nu t}.\label{fneq1:aoup}
\end{align}
The average noise $\langle f\rangle_x$ of the bath particle, for the tracer conditioned at $x$ is given by,
\begin{align}
   \la f\ra_x &= x\frac{\gamma T \Lambda  \tau  (\gamma  \Lambda +\gamma +\Lambda )}{\Lambda  \tau ^2 (\gamma +\Lambda )+\left[\tau  (\gamma  \Lambda +\gamma +\Lambda ) +\gamma \right] (\gamma + \gamma\Lambda T+\Lambda )}.\label{aoup:Af}
\end{align}
This finding leads to a finite $F^{II}(x,t)$ in  Eq.~\eqref{lin:gen}, and using $\la f(t)\ra_x=\la f\ra_x e^{-t/\tau}$,
\begin{align}
    F^{II}(x,t)&=\nu \la f\ra_x
    \int_{0}^t dt' e^{-\nu(t- t')}e^{- t'/\tau}.
    \label{fneq2:aoup}
\end{align}
with $\langle f\rangle_x$ given by Eq.~\eqref{aoup:Af}.
Note that, $F^{II}(x,t)$ brings the time-scale $\tau$ of the bath noise into the relaxation of the tracer.  Equation~\eqref{aoup:Af} shows that conditioning the tracer position yields a finite average of bath noise $f$, which in turn, leads to a finite force on the tracer for $t>0$. This is quite remarkable since the equation of motion of the noise $f$ is independent of $q$ and $x$.

\begin{figure}
    \includegraphics[width=\hsize]{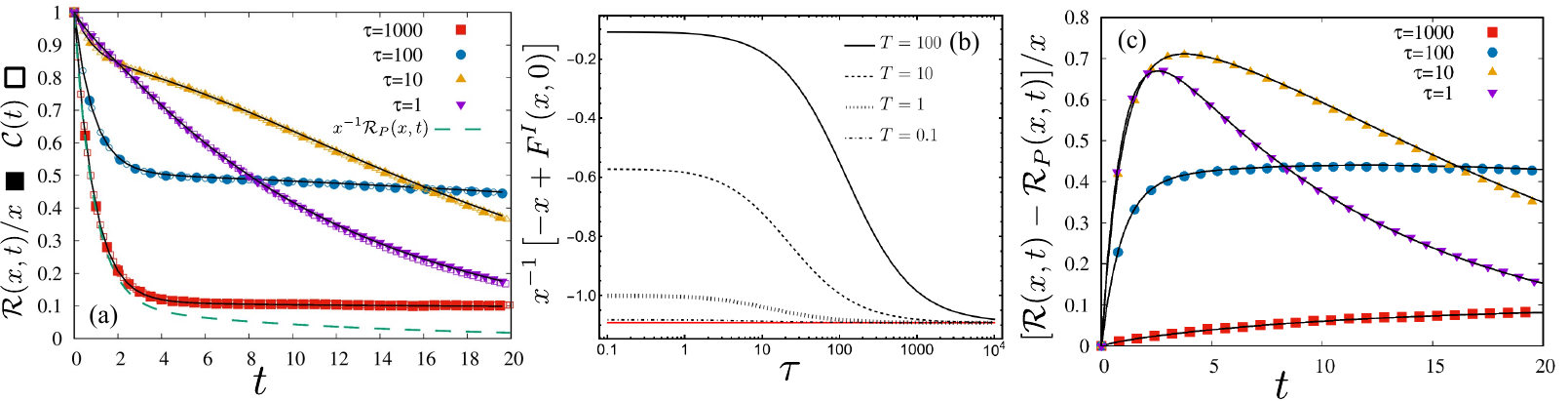}
   \caption{Gaussian active bath: (a) Relaxation function $\mathcal{R}(x,t)/x$ for different values of bath activity; Data from numerical simulations are shown by filled symbols, while the solid black lines correspond to the analytical expression Eq.~\eqref{mcd-ac-linear:th:neq}. The pinned relaxation $\mathcal{R}_P(x,t)/x$ (dashed line), as predicted, is independent of the bath temperature. Open symbols represent  the position autocorrelation $\mathcal C(t)=\la x(t)x)(0)\ra/\la x^2\ra$ for the same set of bath temperatures (see discussion section~\ref{autocorrelation:dis}). (b) shows the magnitude of the initial force on the tracer versus $\tau$ for different values of $T$;  red line gives the saturation value approached for large $\tau$,  Eq.~\eqref{large:tau}. (c) 
   Difference of conditioned and pinned relaxation as a function of time for the same set of bath activities as (a), numerical simulations (filled symbols) and Eq.~\eqref{mcd:rtplin} (solid black lines). For both (a), (b) and (c) $\Lambda=0.1$, and $\gamma=1$; additionally,  $T=100$ for (a) and (c).}
    \label{fig:aoup}
\end{figure}
 The relaxation function, given by Eq.~\eqref{mcd:rtplin}, using Eq.~\eqref{fneq1:aoup} and Eq.~\eqref{fneq2:aoup}, can be obtained as,
\begin{align}
     \mathcal{R}(x,t)=&\mathcal{R}_P(x,t)+ F^{I}(x,0) h^I(t)+\left\la f\right\ra_x h^{II}(t),\label{mcd:rtp:aoup}
 \end{align}
where $\mathcal{R}_P(x,t)$ is given in Eq.~\eqref{eq:pinned:gen}; the functions $ h^{I,II}(t)$ are given in Eq.~\eqref{hI} and Eq.~\eqref{hII}, respectively. 

$\mathcal{R}(x,t)$ is shown in Fig.~\ref{fig:aoup} (a) along with results obtained from numerical simulations. $\mathcal{R}(x,t)$ is a sum of exponentials, and for large times, follows the one corresponding to the largest of the time-scales involved in the functions $h^{I}(t)$ and $h^{II}(t)$, namely, $\max(\tau,(1+\Lambda+\nu))$. Indeed, the curves in Fig.~\ref{fig:aoup}(a) show relaxation with time scale $\tau$  for large values of $\tau$. 

The short-time decay, on the other hand, is as above given via the forces at $t=0$,
\begin{align}
    \frac{d}{dt} \langle x(t)\rangle_{x_0}|_{t=0} =-x_0+F^I(x_0,0)+F^{II}(x_0,0)=-x_0+F^I(x_0,0),\label{eq:meanf:aoup}
    \end{align}
where the second equality follows because  $F^{II}(x,0)=0$. The small time behavior is thus governed by the trap force and $F^{I}(x,0)$. Let us analyze how this force depends on activity $\tau$. For very small activity ($\tau\to 0$),
\begin{align}
     \lim_{\tau\to 0} \frac{F^{I}(x,0)-x}{x}=-\left(1-\frac{\Lambda\gamma(T-1)}{\gamma+\Lambda+\gamma\Lambda T}\right)
\end{align}
which agrees with the result found for different temperatures, see Eq.~\eqref{fneq}.  As is the case for Eq.~\eqref{fneq}, the rhs of the above equation is bound from below by $-x(1+\gamma\Lambda/(\gamma+\Lambda))$, approached for $T\to 0 $,  and bound from above by $0$, approached  for $T\to\infty$. At large activity, i.e., $\tau\to\infty$, the resulting force on the particle saturates to,
\begin{align}
    \lim_{\tau\to\infty} \frac{F^{I}(x,0)-x}{x}=- \left(1+\frac{1}{\gamma^{-1}+\Lambda^{-1}}\right).\label{large:tau}
\end{align}
The force $\left(F^{I}(x,0)-x\right)$ decreases with $\tau$ for intermediate values of $\tau$, and it can be shown to do so monotonically. This is illustrated in Fig~\ref{fig:aoup} (b), and is also evident from the relaxation curves Fig~\ref{fig:aoup} (a), where the short time decay is faster for larger values of $\tau$.

It is also worth noting that in the small activity limit, i.e., $\tau\to 0$, one finds  $\langle f\rangle_x\to 0$, consistent with an effective thermal picture~\cite{santra2022universal} at a temperature $T$. This is also visible in the variance of $f$, Eq.~\eqref{aoup:noisecorr}, which approaches a local version in this limit.

We finish the section by showing that $\mathcal{R}_P(x,t)$, also in this case, corresponds to the physical situation of  the pinned protocol. 
 For pinned $x_0$, the bath dynamics is an active Ornstein-Uhlenbeck process in a harmonic trap centered at $x_0$. Its  Gaussian distribution $P_x(q,f)$ is given by,
\begin{align}
    P_x(q,f)=\frac{1}{2\pi \det \Sigma}e^{-\frac 12 X^T \Sigma^{-1}X}, \text{~with~} X=\begin{pmatrix}
        q-x\\f
    \end{pmatrix},
\end{align}
and the covariance matrix $\Sigma$ given by $\Sigma_{11}=\gamma T/(\Lambda(\gamma+\Lambda\tau)),$ $\Sigma_{22}=\gamma T/\tau$, and $\Sigma_{21}=\Sigma_{12}=\gamma T/(\gamma+\Lambda\tau).$ The pinned marginal distribution of $q$ and $f$ can be easily obtained by respectively integrating over the other variable,
\begin{align}
    P_x(q)=\frac{1}{\sqrt{2\pi\Sigma_{11}}}e^{-\frac{(q-x)^2}{2\Sigma_{11}}},\quad \text{and }P_x(f)=\frac{1}{\sqrt{2\pi\Sigma_{22}}}e^{-\frac{ f^2}{2\Sigma_{22}}}.
\end{align}
From the above pinned distributions, the pinned averages vanish,  
\begin{align}
    \la q-x\ra_{x}^P=0,\quad \text{and  }\la f(t)\ra_{x}^P=0,
\end{align}
where for the last equality we used $\la f\ra^P_x=0$.
This, in turn, implies that the pinned relaxation follows the homogenous part of Eq.~\eqref{aoup:rxt:1}, i.e.,  $\mathcal{R}_P(x,t)$. 


\section{Non-Gaussian active bath}\label{s:rtp}
In this section, we finally consider an active bath particle with non-Gaussian noise, i.e., a run-and-tumble particle (RTP)~\cite{tailleur2008statistical,nash2010run,santra2020run}. The bath noise $f(t)$ is  modeled by a two-state Markov jump process with states $\{a_0,-a_0\}$ [$a_0$ is  force in units of $\sqrt{k k_\text{B}T_{\text{tr}}}$] and (dimensionless) transition rate $\tau^{-1}$. Unlike the previous examples, $f(t)$ is a non-Gaussian noise, characterized by the following  conditioned probability,
\begin{align}
    P(f,t|f_0,0)=\frac{1}{2}\left(1+\frac{ff_0}{a_0^2}e^{-t/\tau}\right).
\end{align}
In the stationary state,  $f$ has the following cumulants, (using $t_4\geq t_3\geq t_2\geq t_1$)
\begin{align}
    \la f(t)\ra&=0\\
    \la f(t)f(t')\ra&=a_0^2 e^{- |t-t'|/\tau}
        \label{rtp:noisecorr}\\
         \la f(t_1)f(t_2)f(t_3)\ra&=0\\
        \la f(t_1)f(t_2)f(t_3)f(t_4)\ra_c&=-2a_0^4\, e^{-(t_4-t_3+t_2-t_1)/\tau}\\
        \vdots\notag
\end{align}
The finiteness of the fourth (and higher)  cumulants demonstrates the non-Gaussianity of the noise. All cumulants decay with time scale $\tau$. 

\begin{figure*}
\centering\includegraphics[width=\hsize]{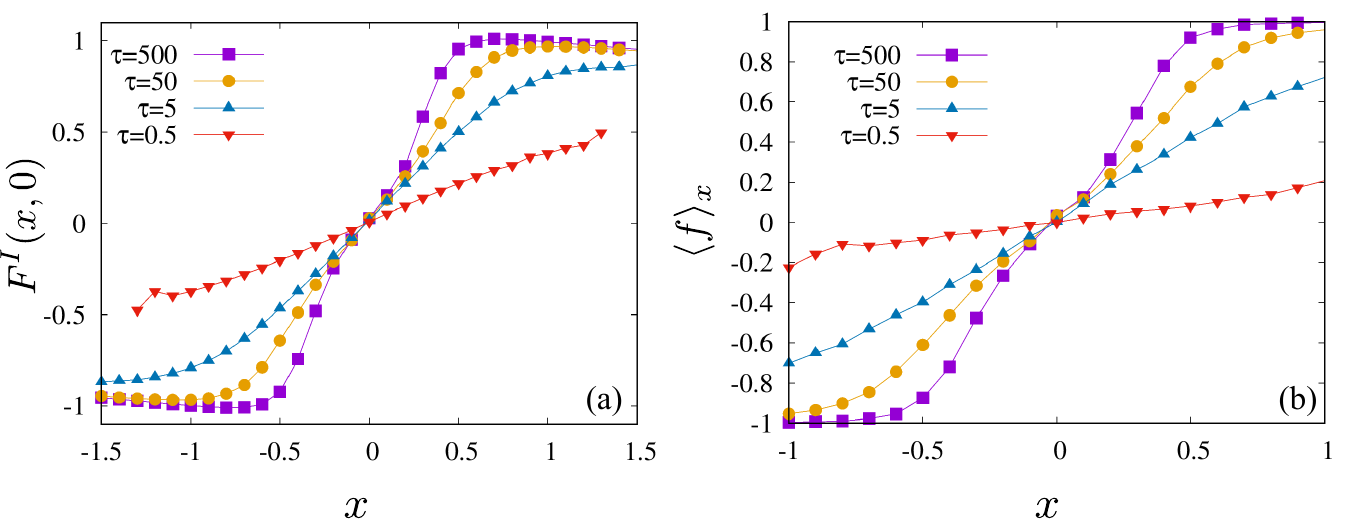}
\caption{Non-Gaussian active bath: (a) shows $F^I(x,0)=\Lambda\la(q-x)\ra_{x}$ for different values of $\tau$. (b) shows the average bath noise $\la f\ra_{x}$ for different values of $\tau$. $k=1$, $\gamma=1$, $\Lambda=0.1$ and $a_0=1$.} 
\label{f:meanoise}
\end{figure*}

\textcolor{black}{Though one can write a Fokker-Planck equation for the joint distribution $P(x,q,f,t)$, solving it appears difficult, even in the stationary state, and we cannot find the non-Gaussian stationary distribution analytically.} 
We thus resort to numerical simulations to find  the conditioned mean of the noise $F(t)$. Fig.~\ref{f:meanoise}(a) shows the resulting conditioned average distance between $x$ and $q$, $\la q_0-x_0\ra_{x_0}$ as a function of $x_0$. Unlike the previous cases, it is a non-linear function of $x_0$ for this model, re-emphasizing the non-Gaussian nature. Figure~\ref{f:meanoise}(b) shows $\la f\ra_{x_0}$, which is also a non-linear function of $x_0$. 
\begin{figure*}
\centering\includegraphics[width=\hsize]{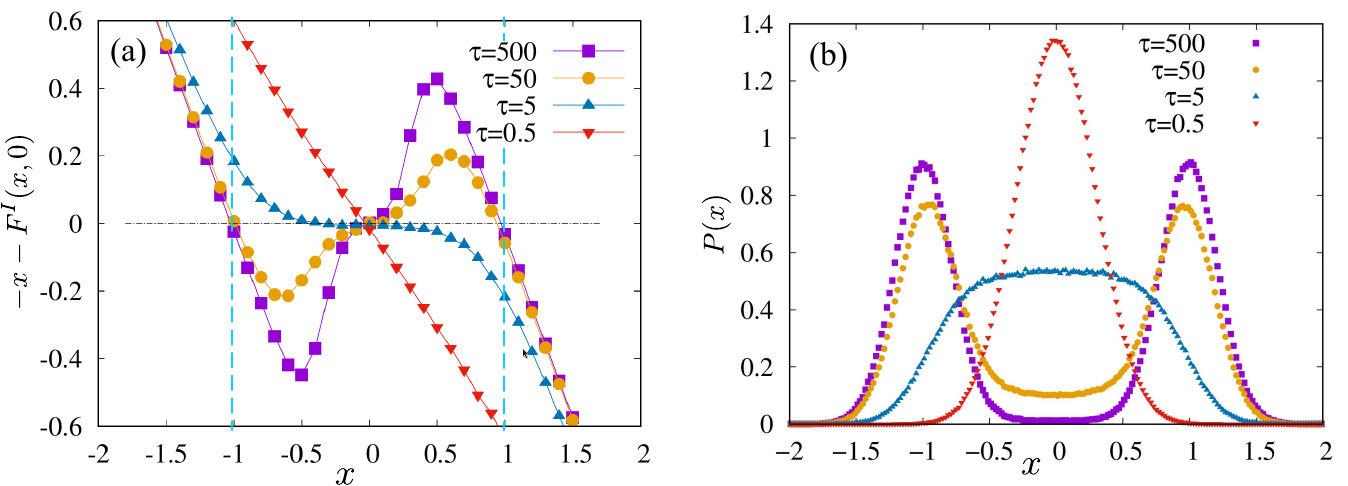}
\caption{Non-Gaussian active bath: (a) shows the force at $t=0$ acting  on the tracer for different values of $\tau$. For large $\tau$,  two zero force points at $\pm x^*=\pm a_0$ (dashed vertical lines) appear, which merge to the origin as $\tau$ decreases (see main text). (b) Stationary position distribution of the tracer;  peaks are nearly at $\pm x^*$, the  force balance points. }
\label{f:netforce-dist}
\end{figure*}

At $t=0$, only the trap force and $F^{I}(x,0)=\Lambda \la q_0-x_0\ra_{x_0}$ are finite, i.e., $F^{II}(x,0)=0$ as before. In Fig~\ref{f:netforce-dist}(a), we plot the net force $-x+\Lambda \la q_0-x_0\ra_{x_0}$. It is, as mentioned, a nonlinear function of $x$, and notably, for some parameter regimes also a non-monotonic one! The non-monotonicity is pronounced for large values of $\tau$, i.e., for large activity, where the graph shows two zero force points, symmetrically on either side of the center of the trap, denoted $\pm x^*$. As activity decreases, these points come closer and coalesce at the origin for small $\tau$. This leads, for large $\tau$, to a bimodal  stationary distribution of $x$,  as seen in Fig~\ref{f:netforce-dist}(b). 

This distribution can be understood in the limit of $\tau\to\infty$, i.e., if $\tau$ is large compared to all other time scales. Then the noise takes the value $\pm a_0$ long enough for the remainder of the system to reach a stationary state under that value, 
 corresponding to the system driven with a time independent force $f=\pm a$. The distribution of $q-x$ is thus Gaussian, centered at  $ (q-x)=\pm a_0\Lambda^{-1}$, where the bath particle at  
$q$ is force free. 
How do this reflect in the tracer distribution $x$? Indeed, in this limit, $x$ also shows a Gaussian, centered at $\pm a_0$, where forces on $x$ are balanced. 
Thus $(x^*,q^*)=(\pm a_0, \pm a_0\pm a_0\Lambda^{-1})$  correspond to zero force points in this limit, which agrees well with  numerical simulations for large $\tau$ in Fig.~\ref{f:netforce-dist} with $a_0=1$.  
The above argument holds for large values of $\tau$, where the active length scale $a_0\tau/\gamma$ is large compared to $x^*+q^*$; for small $\tau$ frequent changes in sign of $f$ does not allow the tracer and bath to reach the stationary Gaussian distributions, and the tracer remains remains confined around $x=0$.

\begin{figure*}
\centering\includegraphics[width=\hsize]{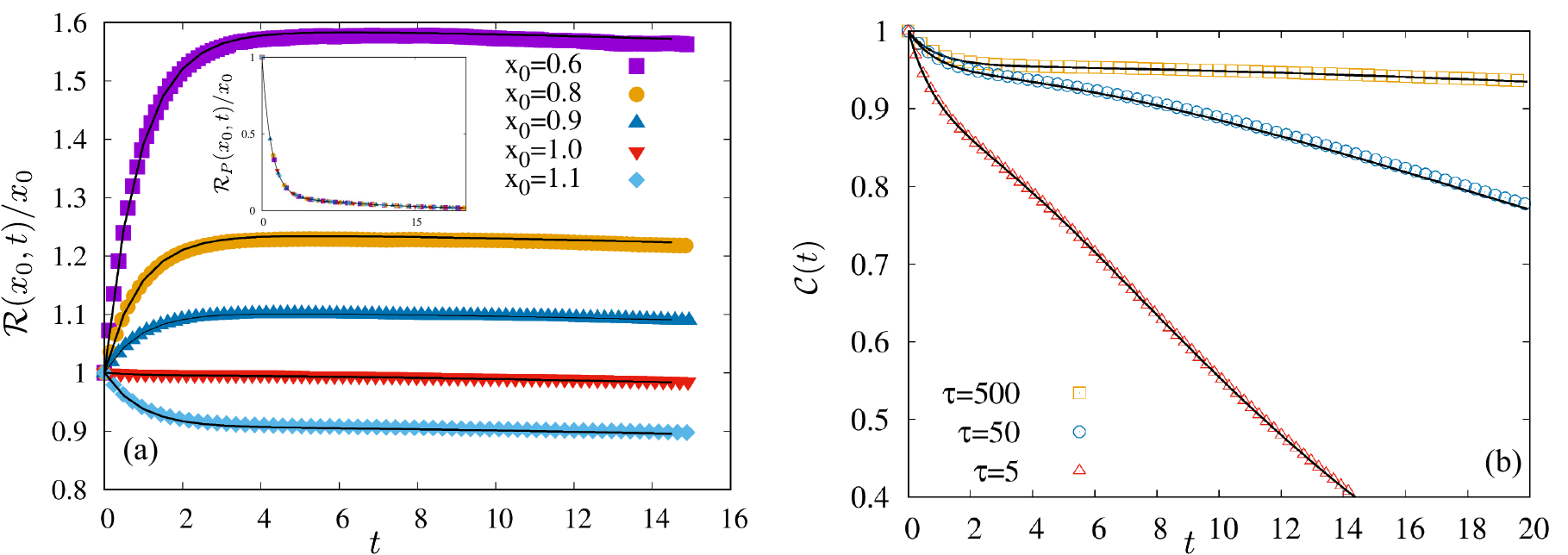}
\caption{Non-Gaussian active bath: (a) $\mathcal{R}(x_0,t)$ for different $x_0$; Data from numerical simulations with $\tau=500$, $k=1$, $\Lambda=0.1$, $\gamma=1$ are shown as symbols, while the solid black lines represent the analytical result Eq.~\eqref{cd:rtp}. For $x^*\lesssim 1$, $\mathcal{R}(x_0,t)$ shows a non-monotonic behavior, with an initial increase from its value at $t=0$; for $x_0\gtrsim x^*=1$, it decays  monotonically. Inset shows the pinned relaxation for the same set of $x_0$. (b) shows the autocorrelation function for different values of $\tau$ (see discussion section~\ref{autocorrelation:dis}): data from numerical simulations are shown with symbols, the analytical forms, Eq.~\eqref{ac}, are shown in solid black lines.}
\label{f:rtp:conditionedrel}
\end{figure*}

As a consequence of this discussion, for $x_0 \lesssim x^*$, the force experienced by the tracer points outwards, i.e., away from the origin. This is reflected by the non-monotonic behavior of the relaxation function for small values of $x_0$ [see Fig~\ref{f:rtp:conditionedrel}]: the initial increase is due to the direction of the net force, and the slow decay at late times is due to the fact that the particle is equally likely to be found around one of the force balance points $\pm x^*$ at long times. On the other hand, for $x_0\gtrsim x^*$, the tracer experiences an inward force, and the conditioned relaxation shows a monotonic decay.
The relaxation function follows
\begin{align}
     \mathcal{R}(x,t)=&\mathcal{R}_P(x,t)+ F^{I}(x,0) h^I(t)+\left\la f\right\ra_x h^{II}(t),\label{cd:rtp}
\end{align}
with the magnitudes of the relative terms being quite different to the previous cases, leading to the nonmonotonic behaviors. Estimating these contributions from Fig.~\ref{f:meanoise}, and using them in Eq.~\eqref{cd:rtp}, yields a consistency check for the conditioned relaxation. This is shown in solid lines, along with numerical simulations in Fig.~\ref{f:rtp:conditionedrel}.

In the case of pinning, the tracer is fixed at $x_0$, and the bath particle follows an equation of motion,
\begin{align}
    \gamma\dot{q}(t)&=-\Lambda(q-x_0)+f(t)\label{eom:bath:pinned},
\end{align}
which is a RTP in a harmonic trap centered at $x_0$. The bath particle reaches a stationary state symmetric about $x_0$~\cite{dhar2019run,fodor2018statistical}. This implies that, $ \la x-q\ra_{x_0}^P=0$. Moreover, in this nonequilibrium stationary state of the bath particle, the noise $f$ is equally likely to take the values $\pm a_0$~\cite{dhar2019run}, implying $\la f\ra _{x}=0$. Thus, also here, the pinned relaxation is the homogeneous solution of Eq.~\eqref{aoup:rxt:1}, ${\mathcal R}_P$, giving the conditioned and the pinned relaxations well defined experimental meanings that allow estimating the nonequilibrium forces $F^{I,II}(x,t)$.



\section{Relaxation and autocorrelation function}\label{autocorrelation:dis}
\textcolor{black}{In the Gaussian cases discussed in this manuscript, $R(x,t)$ is a linear function of $x$. As a consequence, it is directly related to the autocorrelation function, $\mathcal{C}(t)=\la x(t)x(0)\ra/\la x(0)^2\ra$, by
\begin{align}
\mathcal C(t)=\frac{\mathcal{R}(x,t)}{x}.\label{eq:CR}
\end{align}}
One interesting insight from  Eq.~\eqref{eq:CR} is related to a property of $\mathcal C$: In stationary systems, $\mathcal C$ is bound by its initial value, $\mathcal C(t)\leq \mathcal C(0)$ \cite{mclennan1989introduction}. This implies that  in the Gaussian cases considered, $\mathcal{R}(x,t)\leq \mathcal{R}(x,0)$, as we indeed find. 

For the non-Gaussian active case, we observe that $\mathcal{R}(x,t)\leq \mathcal{R}(x,0)$ is not necessarily fulfilled for small values of $x$, while $\mathcal C(t)\leq \mathcal C(0)$ still must hold. No contradiction is implied by the non-monotonic behavior of the conditioned relaxation in Fig.~\ref{f:rtp:conditionedrel}(a), as Eq.~\eqref{eq:CR} is not valid and replaced by a more general relation, 
\begin{align}
    \mathcal{C}(t)=\frac{1}{\la x(0)^2\ra}\int dx P(x) \, x\mathcal{R}(x,t).\label{ac}
\end{align}
This is shown for the run-and-tumble bath particle, along with numerical simulations in Fig.~\ref{f:rtp:conditionedrel}(b).

\section{Conclusion}\label{s:concl}
We investigate the stationary relaxation dynamics of a harmonically confined tracer, an overdamped particle with thermal noise, which is coupled to a bath particle modelled by another overdamped Langevin equation, driven by a noise $f$; the dynamics of $f$ determining the nature of the bath. 
We study three out of equilibrium cases, namely, with $f$ being (i) a thermal noise with a different temperature; (ii) a Gaussian colored noise, emulating an active Ornstein-Uhlenbeck particle, and (iii) $f$  a non-Gaussian colored noise, emulating a run-and-tumble particle. 
We determine the stationary conditioned distributions of the bath degrees of freedom, thereby identifying the forces $F^I(x,t)$ and $F^{II}(x,t)$. $F^I(x,t)$ arises due to the bath degree not being symmetric about the tracer in the stationary conditioned non-equilibrium state, leading to a non-zero tracer bath force. $F^I(x,0)$ and the trap force determine the initial force acting on the tracer and thus the short time decay of the relaxation function. This function is thus, for the non-equilibrium models analyzed,  qualitatively different from equilibrium cases. $F^I(x,0)$ can have either sign, i.e., it can be opposite to the trap force. For the cases (i) and (ii), its magnitude is  however bound by the trap force. For case (iii), $F^I(x,0)$ can exceed the trap force in magnitude, so that the relaxation function can initially increase as a a function of time and  hence be non-monotonic.  
The second force, $F^{II}(x,t)$, is attributed to a non-vanishing mean of $f$ under the conditioning, and it arises for the active baths studied in cases (ii) and (iii).  $F^{II}(x,t)$ vanishes at $t=0$ but influences the relaxation dynamics at later times. 


A  pinning protocol yields a relaxation function which is not influenced by mentioned forces, so that we expect an experimental comparison of the two types of relaxation functions to yield insightful information about the found forces. \textcolor{black}{More specifically, in the model studied,  we find conditional and pinned relaxations to be identical in equilibrium (in \ref{apendix:eq} we argue that this equality holds also for nonlinear potentials in equilibrium), and a difference of the two signals a non-equilibrium process. This conclusion does not contradict the statement that a 1d stationary Gaussian process obeys detailed balance and can thus not be distinguished from an equilibrium system \cite{mitterwallner2020non,klimek2024data,netz2310multi, muenker2022onsager}: The proposed pinning experiment is distinct from the stationary one, and thus yields additional information.}
Pinning may, e.g., be experimentally achieved via optical forces \cite{volpe2023roadmap}. Indeed, a variety of non-equilibrium baths have been realized, including tracers suspended in swimmers~\cite{wu2000particle,park2020rapid,boriskovsky2024fluctuation}. 

Future work can investigate the properties of the relaxation function in systems driven out of equilibrium by oscillating drives~\cite{cui2018generalized} or shear flows~\cite{pelargonio2023generalized}. 
It can also be interesting to investigate possible implications of the non-equilibrium forces on non-equilibrium transport~\cite{santra2022activity,sarkar2024harmonic} and thermodynamic quantities~\cite{guevara2023brownian,ginot2023average}. Recently, a quantity called mean back relaxation has been introduced in ~\cite{muenker2022onsager,knotz2023mean}. Investigating if mean back relaxation can be obtained from a Langevin description with the non-zero conditioned means is an open question. Models of tracers coupled to fluctuating fields~\cite{gross2021dynamics,basu2022dynamics,demery2019driven} have been studied lately, revealing power-law decays, memory-induced oscillations in transient relaxation and \textcolor{black}{interesting thermodynamic properties}~\cite{venturelli2022nonequilibrium,venturelli2023memory,venturelli2024stochastic}; studying the stationary relaxation of tracers coupled to nonequilibrium fields will also be an interesting direction.

\section{Acknowledgements}
The authors would like to thank Urna Basu, Christian Maes, and Juliana Caspers for discussions.

\appendix

\section{Computation of position distribution for the case of tracer and bath at different temperatures}\label{app:neqdist}
In this appendix, we derive some of the results used for the the case of tracer and bath at different temperatures. Since the set of equations are coupled linear equations with Gaussian white noises, the joint stationary distribution of the position of the tracer and bath particle should thus be a bivariate Gaussian, of the form,
\begin{align}
    P(x,q)=\frac{1}{2\pi\sqrt{\det \Sigma}}\exp[-X^T\Sigma^{-1}X],
\end{align}
where $X^T=\Big[x~~~q \Big]$ and $\Sigma=\begin{bmatrix}
\la x^2\ra & \la xq \ra\\
\la qx\ra & \la q^2\ra 
\end{bmatrix}$
denotes the covariance matrix. The elements of the covariance matrix can be solved by solving the set of equations in the Fourier domain,
\begin{align}
    \tilde x(\omega)&=\frac{\Lambda\tilde f(\omega)+\tilde{\eta}(\omega)(\Lambda-i\gamma\omega)}{(\omega^2\gamma-\Lambda)+i\omega(\gamma+\Lambda+\Lambda\gamma)}\\
    \tilde q(\omega)&=\frac{\Lambda\tilde\eta(\omega)+\tilde f(\omega)((1+\Lambda)-i\omega)}{(\omega^2\gamma-\Lambda)+i\omega(\gamma+\Lambda+\Lambda\gamma)}\cr
\end{align}
Using the correlations of the noises in the Fourier space, i.e.,
$\la\tilde\eta(\omega)\tilde\eta(\omega')\ra=2\pi\delta(\omega+\omega')$, $\la\tilde f(\omega)\tilde f(\omega')\ra=4\pi D_b\delta(\omega+\omega')$ and $\la\tilde\eta(\omega)\tilde f(\omega')\ra=0$, we have,
\begin{align}
    \la x^2\ra &=\frac{\Lambda^2(\gamma T+1)+\gamma^2\lambda_1\lambda_2}{\lambda_1\lambda_2 (\lambda_1+\lambda_2)},\\
    \la q^2\ra &=\frac{\Lambda^2+\gamma T((1+\Lambda)^2+\lambda_1\lambda_2)}{\lambda_1\lambda_2 (\lambda_1+\lambda_2)},\\
    \la x q\ra&=\frac{\Lambda^2+\gamma T\Lambda(1+\Lambda)}{\lambda_1\lambda_2 (\lambda_1+\lambda_2)},
\end{align}
with $\lambda_1$ and $\lambda_2$ being given by,
\begin{subequations}
\begin{align}
    \lambda_1=&\frac{1}{2\gamma}(\gamma+\Lambda+\gamma\Lambda+\sqrt{(\gamma+\Lambda+\gamma\Lambda)^2-4\gamma\Lambda})\\
    \lambda_2=&\frac{1}{2\gamma}(\gamma+\Lambda+\gamma\Lambda-\sqrt{(\gamma+\Lambda+\gamma\Lambda)^2-4\gamma\Lambda})
\end{align}
\label{betagamma}
\end{subequations}

To obtain the force experienced by the tracer [in Eq.~\eqref{eff:noise:eq}] we need to compute $\la x-q\ra_{x_0}$. To this end, we compute the conditioned distribution $P(q|x_0)$,
\begin{align}
    P(q|x_0)&=\mathcal{Z}^{-1}\int_{-\infty}^{\infty}dx~\delta(x-x_0)P(x,q)\cr
    &=\frac{1}{\sqrt{4\pi (\la q^2\ra-\la xq\ra^2/\la x^2\ra\ra)}}\exp\left[-\frac{\left(q-x_0\frac{\la xq\ra}{\la x^2\ra}\right)^2}{2(\la q^2\ra-\la xq\ra^2/\la x^2\ra\ra)} \right].
\end{align}
which leads to, 
\begin{align}
    F^{I}(x_0,0)=-\Lambda(x_0-\la q\ra_{x_0})=-\Lambda x_0\left(1-\frac{\la xq\ra}{\la x^2\ra}\right).
\end{align}

 Additionally, the two-point correlation $\la F(t);F(t')\ra=\la F(t)F(t')\ra-\la F(t)\ra \la F(t')\ra$  is given by,
 \begin{align}
   \la F(t);F(t')\ra_{x_0}=&2\delta(t-t')+\Lambda T e^{-\Lambda (t-t')}+\Lambda e^{-\Lambda(t+t')}\left[\Lambda \la q;q\ra_{x_0}-T\right]\\
     =& 2\delta(t-t')+\Lambda T e^{-\Lambda (t-t')}+\Lambda e^{-\Lambda(t+t')}\left[\frac{\Lambda ^2 (T+1)^2+4 \Lambda  T+T}{(2 \Lambda +1) (\Lambda +\Lambda  T+1)}-T\right].\nonumber
     \end{align}
  Note that the final term in the above equation is transient. For the equilibrium case ($T=1$), it becomes zero and we recover FDT. However, for nonequilibrium cases, the transient term vanishes only for large $t,t'$. This is suggestive of the conditioned two point autocorrelation for the nonequilibrium cases to be also non-trivial. It would be interesting to see if the mean back relaxation~\cite{knotz2023mean} can be computed from using the above conditioned correlation.

\section{Computation of position distribution for the bath of Gaussian active particles}\label{app:aoup}
Here, we consider the case of a bath of active-Ornstein-Uhlenbeck particle
\begin{align}
    \dot{x}(t)&=-x-\Lambda(x-q)+\eta(t)\label{eom:tracer1:aoup}\\
\gamma\dot{q}(t)&=-\Lambda(q-x)+f(t)\label{eom:bath1:aoup},\\
\tau\dot{f}(t)&=-f(t)+\xi(t),\label{noise:aoup:appendix}
\end{align}
where the noise correlations are given by,
\begin{align}
    \la f(t)\ra_{f_0}=f_0e^{ -t/\tau}\quad \la f(t)f(t')\ra_{f_0}=\frac{\gamma T}{\tau} e^{- |t-t'|/\tau}.
    \label{aoup:noisecorr:appendix}
\end{align}
Since the equations of motion are linear in the Gaussian noises, the joint distribution $P(x,q,f)$ is a trivariate Gaussian,
\begin{align}
    P(x,q,f)=\frac{1}{2\pi\sqrt{\det \Sigma}}\exp[-X^T\Sigma^{-1}X],
\end{align}
where $X^T=\begin{pmatrix}
    x~~~q~~f
\end{pmatrix}$ and $\Sigma=\begin{pmatrix}
\la x^2\ra & \la xq \ra & \la xf\ra\\
\la qx\ra & \la q^2\ra & \la qf\ra\\
\la xf\ra & \la qf & \la f^2\ra
\end{pmatrix}$. The elements of the covariance matrix can be obtained by taking a Fourier transform of the coupled equations in $(x,q,f)$ in the same manner as the two temperature case. Below we provide only the elements required to compute the forces, \eqref{lin:gen}
\begin{align}
    \la x^2\ra=&\frac{\gamma ^2 \lambda _1 \lambda _2+\Lambda ^2}{\lambda _1 \lambda _2 \left(\lambda _1+\lambda _2\right)}+\gamma T\frac{\Lambda ^2  \left(\left(\lambda _1+\lambda _2\right) \tau +1\right)}{\lambda _1 \lambda _2 \left(\lambda _1+\lambda _2\right) \left(\lambda _1 \tau +1\right) \left(\lambda _2 \tau +1\right)},\\
    \la q^2\ra=&\frac{\Lambda^2}{\lambda_1\lambda_2(\lambda_1+\lambda_2)}+\gamma T\frac{\lambda _2 \tau +\lambda _1 \left(\lambda _2+\tau \right)+1}{\lambda _1 \lambda _2 \left(\lambda _1+\lambda _2\right) \left(\lambda _1 \tau +1\right) \left(\lambda _2 \tau +1\right)}\\
    \la f^2\ra =& \frac{\gamma T}\tau\\
    \la xq\ra =&\frac{\Lambda ^2}{\lambda _1 \lambda _2 \left(\lambda _1+\lambda _2\right)}+\gamma T\frac{\left(\Lambda ^2+\Lambda \right) \left(\left(\lambda _1+\lambda _2\right) \tau +1\right)}{\lambda _1 \lambda _2 \left(\lambda _1+\lambda _2\right) \left(\lambda _1 \tau +1\right) \left(\lambda _2 \tau +1\right)},\\
    \la xf \ra =&\frac{ \gamma T\,\Lambda \left(\lambda _1 \left(\Lambda  \tau -\gamma  \lambda _2\right)+\lambda _2 \Lambda  \tau +\Lambda \right)}{\lambda _1 \lambda _2 \left(\lambda _1+\lambda _2\right) \left(\lambda _1 \tau +1\right) \left(\lambda _2 \tau +1\right)},\\
\end{align}
where $\lambda_1$ and $\lambda_2$ are still given by Eq.~\eqref{betagamma}.

The conditioned stationary distribution of the tracer, given a position $x$ of the tracer is given by,
\begin{align}
    P(q|x_0)&=\frac{1}{\sqrt{4\pi (\la q^2\ra-\la xq\ra^2/\la x^2\ra\ra)}}\exp\left[-\frac{\left(q-x_0\frac{\la xq\ra}{\la x^2\ra}\right)^2}{2(\la q^2\ra-\la xq\ra^2/\la x^2\ra\ra)} \right].
\end{align}
This leads to the force arising due to the bath position,
\begin{align}
    F^{I}(x_0,0)=&-\Lambda(x_0-\la q\ra_{x_0})=-\Lambda x_0\left(1-\frac{\la xq\ra}{\la x^2\ra}\right)\cr
    &=-\Lambda x_0\frac{\gamma  \Lambda  \tau ^2+(\gamma -1) (\tau  (\gamma  \Lambda +\gamma +\Lambda )+\gamma )}{\Lambda  \tau ^2 (\gamma +\Lambda )+(\gamma +2 \Lambda ) (\tau  (\gamma  \Lambda +\gamma +\Lambda )+\gamma )}.\label{x-q:aoup}
\end{align}
Similarly,
\begin{align}
    P(f|x_0)&=\frac{1}{\sqrt{4\pi (\la f^2\ra-\la xf\ra^2/\la x^2\ra\ra)}}\exp\left[-\frac{\left(f-x_0\frac{\la xf\ra}{\la x^2\ra}\right)^2}{2(\la x^2\ra-\la xf\ra^2/\la x^2\ra\ra)} \right].
\end{align}
which leads to,
\begin{align}
    f_{x}&=x\frac{\Lambda  \tau  (\gamma  \Lambda +\gamma +\Lambda )}{\Lambda  \tau ^2 (\gamma +\Lambda )+(\gamma +2 \Lambda ) (\tau  (\gamma  \Lambda +\gamma +\Lambda )+\gamma )}.\label{f:aoup}
\end{align}

{\color{black}
\section{Equivalence of conditioned and pinned relaxation in equilibrium}\label{apendix:eq}
In the main text we find that conditioned and pinned relaxation are identical in equilibrium, for the linear system investigated. In this Appendix, based on the Boltzmann distribution, we demonstrate that this equality is more general, i.e., it is valid for nonlinear interactions and nonlinear confining potentials $U(x)$. Let us consider that the tracer degree of freedom ($x$) is coupled to multiple bath degrees of freedom $(\{q_i(t)\})$ via an interaction potential $V(\{\Delta_i\}=\{x-q_i\})$. The joint distribution, given by the Boltzmann distribution (taking $k_\text{B}T=1$),
\begin{align}
    P(x,\{\Delta_i\})=\frac{e^{-[V(\{\Delta_i\})+U(x)]}}{\displaystyle\int dx e^{-U(x)}\displaystyle\int \displaystyle\prod_i d\Delta_i e^{-V(\{\Delta_i\})}},
\end{align}
is in a product form $ P(x,\{\Delta_i\})=P(x)P(\{\Delta_i\})$. The conditioned distribution of the bath, given the tracer position, is thus given by
\begin{align}
    P(\{\Delta_i\}|x)=\frac{e^{-V(\{\Delta_i\})}}{\displaystyle\int \displaystyle\prod_i d\Delta_i  e^{-[V(\{\Delta_i\})]}}.\label{conditioned:dist:gen}
\end{align}
For pinning, on the other hand, the bath particles attain the equilibrium distribution about the pinned position $x$ of the tracer, which is,
\begin{align}
    P_x(\{\Delta_i\})=\frac{e^{-V(\{\Delta_i\})}}{\displaystyle\int \displaystyle\prod_i d\Delta_i  e^{-[V(\{\Delta_i\})]}}.
\end{align}
which is equal to Eq.~\eqref{conditioned:dist:gen}. Thus, in equilibrium, the initial distribution for the pinned and conditioned relaxations are the same, implying that these relaxations are the same for all time $t$. This finding is thus not restricted to linear systems. 
}

\centering---------------------------------

\end{document}


\section{Appendix}
\subsection{Equation of motion for the autocorrelation functions for a Hamiltonian system}
\textcolor{blue}{I would prefer to introduce here the correlation function via the joint distribution $W_2$ (for example sec 1.5 in my thesis). The reason is that the approach with the "Heisenberg picture" works in some cases but maybe not all (I think it does not work for FP equation).} 
\textcolor{blue}{I would prefer to not write the equation for $A$. Also, I think $A$ has an implicit time dependence through $X$, not an explicit one. }  

In this section, we discuss that the generalized Langevin equation obtanied using Zwanzig-Mori projection operator formalism predicts the correlation functions correctly. Let us consider a system governed by a time independent Hamiltonian $H(X)$ where $X$ is a vector comprising of the position $x$ and momentum $p$ of the system [we consider a one-dimensional system for notational convenience, the generalization to higher dimensions is trivial]. Using Hamilton's equations of motion, the time evolution of the phase space distribution function $P(X,t)$ can be shown to follow,
\begin{align}
    \frac{\partial P( X,t)}{\partial t}=-\mathbb{L}P(X,0)
\end{align}
where $\mathbb{L}=\left(\partial_p H\cdot \partial_x-\partial_x H\cdot\partial_p\right)$ denotes the Liouville operator. The general soltuion to the above equation are given by,
\begin{align}
   P(X,t)=e^{-\mathbb{L}t}P(X,0),
\end{align}

The autocorrelation function $\mathcal{C}(t)$ of the variable $A(X,t)$ is definted by,
\begin{align}
    \mathcal{C}(t)=\int dX \int dX' W_2(Xt,X'0) P(X',0) A(X)A(X'),
\end{align}
where $W_2(Xt,X'0)=e^{-\mathbb L t}\delta(X-X')$ denotes the joint probability that the system was at $X'$ at $t=0$ and evolves to a state $X$ at a later time $t$. Using the fact that, $\mathbb L$ is an anti self-adjoint operator, one can rewrite the above equation as,
\begin{align}
    \mathcal{C}(t)=&\int dX P(X) A(X)e^{\mathbb L t}A(X)\cr
    &=\la A(X)e^{\mathbb L t}A(X)\ra.
    \label{defn:avg}
\end{align}

To obtain an equation of motion for $\mathcal{C}(t)$, we take a time derivative of the above equation (for notational convenience, we drop the argument $X$ in the following discussion),
\begin{align}
    \frac{\partial{\mathcal{C}}(t)}{\partial t}=\frac{\partial}{\partial t}\la A e^{\mathbb{L}t}A\ra=\la Ae^{\mathbb{L}t}\,\mathbb{L} A\ra.
\end{align}
We now define a projection operator $\mathbb P$ to the subsapce of $A$, which is defined as $PZ=\la Z\,A\ra/\la A\,A\ra\,A$. Thus, the Liouville operator $\mathbb L$ can be decomposed into two orthogal parts, $\mathbb {PL}$ and $(1-\mathbb P)\mathbb L$. Using this, we can rewrite the above equation as,
\begin{align}
     \frac{\partial{\mathcal{C}}(t)}{\partial t}=\la A e^{\mathbb L t}(1-\mathbb P) \mathbb L \ra-k \mathcal{C}(t),
\end{align}
where $k=-\la \mathbb L A\,A\ra/C(0)$. We now need to simplify the first term on the rhs of the above equation, which we call $I(t)$ for simplicity. Using the Dyson's identity,
\begin{align}
    I(t)&=\la A\left[ e^{(1-\mathbb P)\mathbb Lt}(1-\mathbb P)\mathbb L \right] A\ra\cr
    &+\la A\int_0^t dt' e^{\mathbb L(t-t')}\mathbb {PL}e^{(1-\mathbb P)\mathbb L t'}(1-\mathbb P)\mathbb LA\ra.~~~\label{I(t)}
\end{align}
Note that, in the first term, the operator inside square brackets projects everything to the right of itself to a space orthogonal to $A$ and thus the inner product vanishes. Thereafter using the definition of the projection operator, we have,
\begin{align}
    I(t)=\la A\int_0^t dt' e^{\mathbb L(t-t')}\frac{\mathbb Le^{(1-\mathbb P)\mathbb Lt'}(1-\mathbb P)\mathbb LA\, A}{\mathcal{C}(0)} A\ra.
\end{align}
Thus,
\begin{align}
    \frac{\partial{\mathcal{C}}(t)}{\partial t}=-k\mathcal{C}(t)-\int_0^t dt'\Gamma(t-t')\mathcal{C}(t'),
\end{align}
where $\Gamma(t)=\mathbb Le^{(1-\mathbb P)\mathbb Lt'}(1-\mathbb P)\mathbb LA\, A/\mathcal{C}(0)$.
Taking  a Laplace transform, we get the autocorrelation in Laplace space as,
\begin{align}
    \tilde {\mathcal{C}}(s)=\frac{\mathcal{C}(0)}{s+k+\tilde\Gamma(s)}
    \label{autocorr:eq}
\end{align}
The same result can be obtained for the time evolution of the autocorrelations starting from a Zwanzig-Mori generalized Langevin equation,
\begin{align}
    \frac{\partial{A}(t)}{\partial t}=-kA(t)-\int_0^t dt'\Gamma(t-t')A(t')+F(t)
\end{align}
where $F(t)$ is the noise term [see section xxx of Ref.~\cite{}]. Thus for equilibrium dynamics, a linear Generalized Langevin equation predicts the autocorrelation function correctly.

Let us now consider the average of the observable $A(X,t)$, with the condition that $A(X,0)=A_0$. The conditioned average is defined by,
\begin{align}
    \mathcal{M}(t)=& \int dX'\int dX \delta (A(X')-A_0)W_2(Xt,X'0)\cr
    &\times P(X',0) A(X) ~~   =\la\delta(A-A_0) e^{\mathbb Lt} A\ra,
\end{align}
where $\la\cdots\ra$ has been defined in \eqref{defn:avg}.

Proceeding similarly, as for the $\mathcal{C}(t)$ calculation, we can write,
\begin{align}
     \frac{\partial{\mathcal{M}}(t)}{\partial t}=\left\la \delta(A-A_0)e^{\mathbb L t}(1-\mathbb P) \mathbb L A\right\ra-k\, \mathcal{M}(t),
\end{align}
The first term in the above equation, lets call it $J(t)$, can be  written following Dyson's identity as,
\begin{align}
    J(t)&=\left\la \delta(A-A_0)\left[ e^{(1-\mathbb P)\mathbb Lt}(1-\mathbb P)\mathbb L \right] A\right\ra\cr
    &+\left\la \int_0^t dt' e^{\mathbb L(t-t')}\mathbb {PL}e^{(1-\mathbb P)\mathbb L t'}(1-\mathbb P)\mathbb LA\right\ra.~~~\label{J(t)}
\end{align}
Unlike the case of autocorrelations, here the first term is not in general zero and is a function of $A_0$ and $t$, say $F(A_0,t)$. The second term can, however, be treated in a similar manner and we finally get,
\begin{align}
    \frac{\partial{\mathcal{M}}(t)}{\partial t}=-k\mathcal{M}(t)&-\int_0^t dt'\Gamma(t-t')\mathcal{M}(t')+F(A_0,t).
\end{align}
The mean conditioned average is obtained by integrating over the initial value $A_0$ with its equilibrium weight $p(A_0)$,
\begin{align}
   \mathcal{M}_m(t)=\int dA_0 p(A_0) \la\delta(A-A_0) e^{\mathbb Lt} A\ra
\end{align}
leading to 
\begin{align}
    \frac{\partial \mathcal{M}_m(t)}{\partial t}=-k\mathcal{M}_m(t)&-\int_0^t dt'\Gamma(t-t')\mathcal{M}_m(t')+\bar F(t),
\end{align}
where $\bar F(t)=\int dA_0 p(A_0)F(A_0,t)$. Thus, the mean conditioned average and autocorrelation $\mathcal{C}(t)$ follow different equations of motion, even in equilibrium. In particular, the equation of motion of the MCD has an additional term which depends explicitly on the conditioning. In fact, an arbitrary autocorrelation function $\mathcal{C}_{AB}(t)=\la B(0)A(t)\ra$ follows,
\begin{align}
    \frac{\partial \mathcal{C}_{AB}(t)}{\partial t}=-k \mathcal{C}_{AB}(t)&-\int_0^t dt'\Gamma(t-t') \mathcal{C}_{AB}(t')+\bar F_{AB}(t),
\end{align}
where $k$ and $\Gamma$ are still given as before, and $\bar F_{AB}(t)=\la B\left[ e^{(1-\mathbb P)\mathbb Lt}(1-\mathbb P)\mathbb L \right] A\ra$.

\subsection{Derivation of the generalized Langevin Eq.~\eqref{linear:exactgle} for the tracer}
For the tracer and bath interacting via harmonic springs of stiffness $\Lambda$, the equations of motion,
\begin{align}
\dot{x}(t)&=-kx-\Lambda(x-q)+\sqrt{2T}\eta(t)\label{eom:tracer1APP}\\
\gamma\dot{q}(t)&=-\Lambda(q-x)+f(t)\label{eom:bath1APP}.
\end{align}
The general solution of the bath equation can be written as,
\begin{align}
    q(t)=q_0
\end{align}


\begin{figure}
\includegraphics[width=0.5\hsize]{largetb.pdf}
\caption{MCD for different values of bath temperatures with $k=1$, $V_0=0.1$, $l_0=10$, $\gamma=1$, $T=0.001$. For very large bath temperatures the tracer behaves like an Ornstein-Uhlenbeck particle in a trap where MCD decays as $e^{-kt}$ [denoted by the solid black line].}
\label{f:largeTB}
\end{figure}

\bibliography{ref}